\documentclass[preprint,aip,apl,showpacs,amsfonts,amssymb,superscriptaddress]{revtex4}
\usepackage{stmaryrd}
\usepackage{graphicx}
\usepackage{stmaryrd}
\usepackage{cellspace}

\begin{document}
\title{Transition from strong to ultra-strong coupling regime in mid-infrared metal-dielectric-metal cavities}
\author{P. Jouy}
\affiliation{Laboratoire ``Mat\'eriaux et Ph\'enom\`enes Quantiques'', Universit\'e Paris Diderot, CNRS-UMR 7162, 75013 Paris, France}
\author{A. Vasanelli}
\email{angela.vasanelli@univ-paris-diderot.fr}
\affiliation{Laboratoire ``Mat\'eriaux et Ph\'enom\`enes Quantiques'', Universit\'e Paris Diderot, CNRS-UMR 7162, 75013 Paris, France}
\author{Y. Todorov}
\affiliation{Laboratoire ``Mat\'eriaux et Ph\'enom\`enes Quantiques'', Universit\'e Paris Diderot, CNRS-UMR 7162, 75013 Paris, France}
\author{A. Delteil}
\affiliation{Laboratoire ``Mat\'eriaux et Ph\'enom\`enes Quantiques'', Universit\'e Paris Diderot, CNRS-UMR 7162, 75013 Paris, France}
\author{G. Biasiol}
\affiliation{IOM CNR, Laboratorio TASC, Area Science Park, I-34149 Trieste, Italy}
\author{L. Sorba}
\affiliation{NEST, Istituto Nanoscienze-CNR and Scuola Normale Superiore, I-56127 Pisa, Italy}
\author{C. Sirtori}
\affiliation{Laboratoire ``Mat\'eriaux et Ph\'enom\`enes Quantiques'', Universit\'e Paris Diderot, CNRS-UMR 7162, 75013 Paris, France}


\begin{abstract}
We have investigated the transition from strong to ultra-strong coupling regime between a mid-infrared intersubband excitation and the fundamental mode of a metal-dielectric-metal microcavity. The ultra-strong coupling regime is demonstrated up to room temperature for a wavelength of $11.7~\mu$m by using 260~nm thick cavities, which impose an extreme sub-wavelength confinement. By varying the doping of our structures we show that the experimental signature of the transition to the ultra-strong coupling regime is the opening of a photonic gap in the polariton dispersion. The width of this gap depends quadratically on the ratio between the Rabi and intersubband transition energies.
\end{abstract}

\maketitle
Intersubband polaritons are quasi-particles issued from the coupling between an intersubband (ISB) excitation of a two-dimensional
electron gas and a microcavity photon mode~\cite{dini}. In the mid infrared region, ISB polaritons have been demonstrated with planar
 waveguides~\cite{dini,sapienza,dupont}, by coupling the ISB transition with a continuum of photonic states.
 Using these systems, electroluminescent devices operating in the strong coupling regime up to room temperature have been
 realized~\cite{sapienza}. In the THz frequency range, the strong coupling regime has been achieved by coupling the ISB transition
 with the fundamental mode of a metal dielectric metal (MDM) microcavity~\cite{todorov_PRL2009}. These cavities are widely used
 to realize mid and far infrared quantum cascade lasers~\cite{unterrainer, williams}. This system allows a highly
 sub-wavelength confinement of the light as the thickness of the cavity, $L$, is a small fraction of the wavelength
 ($\lambda/L \approx 1/100$)~\cite{hibbins_PRL2004}. The coupling energy $2\hbar \Omega_R$, which is inversely proportional to $\sqrt{L}$,
 can be therefore strongly enhanced and made comparable with the ISB transition energy $E_{12}$. Recently,~\cite{todorov_PRL2010}
 the record value $2\hbar \Omega_R/E_{12}=0.48$ has been achieved for a transition frequency of 3~THz. In this case the system
 enters the so-called ultra-strong coupling regime~\cite{hubers, ciuti_PRB2005}, characterised by the fact that the anti-resonant
 and the quadratic terms of the light-matter interaction cannot be neglected. An experimental signature of this regime is the
 opening of a photonic gap in the polariton dispersion. In this work, we demonstrate that MDM cavities can be used also in the
 mid infrared wavelength range (5 -15~$\mu$m) to achieve the ultra-strong coupling regime. By studying three samples, based
 on identical quantum wells but with a different doping level, we have investigated the transition from the strong to the
 ultra-strong coupling regime, up to room temperature.

\begin{figure}[ht]
\centering
\includegraphics[width=0.8\columnwidth]{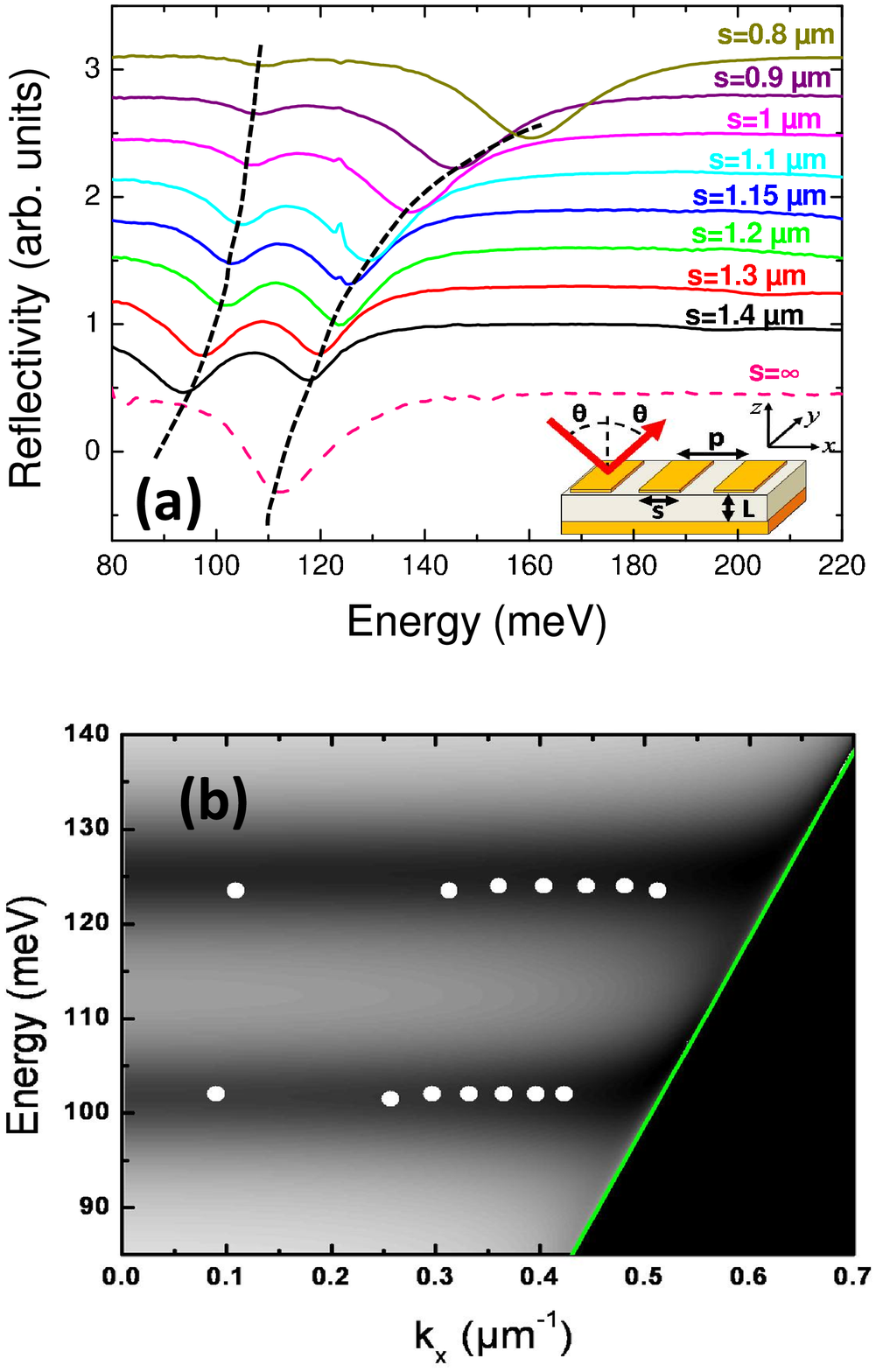}
\caption{(a) Reflectivity spectra measured for sample HM2952 at 300~K for different values of the strip width. The period $p$ is $1.5~\mu$m. The spectrum in dashed line presents the transmission measured on an unprocessed sample in the multi-pass geometry. The inset presents a scheme of our devices, together with the relevant parameters of the grating. (b) Reflectivity minima (white dots) and simulated reflectivity (grey scale) as functions of $k_x$ for $s=1.2~\mu$m; the line represents the light cone.}
\label{reflectivity}
\end{figure}

Our system is composed of a $L = 259$ nm thick semiconductor layer, containing 10 doped GaAs/Al$_{0.35}$Ga$_{0.65}$As quantum wells (9~nm/14~nm), sandwiched between two golden plates, one of which is patterned as a strip grating (see inset of figure~\ref{reflectivity}a). The MDM structure supports photonic modes, whose energy can be tuned by changing the width, $s$, of the metallic strips~\cite{jouy_apl}. The grating period $p$, which in our structures is equal to $1.5~\mu$m, determines the coupling efficiency with the outside radiation. The energy difference between the first two energy levels is $E_{12} = 106$~meV. In order to study the strength of the coupling, we realized three samples (HM2892, HM2951, HM2952) with the same quantum well but with different doping levels. Samples HM2892 and HM2951 are delta doped in the middle of the barriers, while for HM2952 the Si doping extends in the barriers for a length of 12~nm (1~nm of AlGaAs at each side of the quantum wells is left undoped). Shubnikov-de Haas measurements were performed at 1.5~K, resulting in the following values of the electronic densities: $8 \times 10^{11} \, \textrm{cm}^{-2}$ (HM2892), $1.5 \times 10^{12} \, \textrm{cm}^{-2}$ (HM2951), $3 \times 10^{12} \, \textrm{cm}^{-2}$ (HM2952).

The samples were characterized by measuring reflectivity spectra at an angle $\theta = 10^\circ$. The spectra obtained at room temperature for sample HM2952 are shown in Fig. 1a, for different values of the strip width. Two minima are observed in each spectrum, corresponding to the upper and lower polariton states. Their energy position varies with $s$, as this parameter determines the detuning between the cavity mode and the ISB transition energy. In the figure is also illustrated by a dashed line the transmission spectrum of the quantum well measured on an unprocessed sample at room temperature in the multi-pass geometry. A minimum is observed, corresponding to the ISB absorption at energy $\tilde{E}_{12}=\sqrt{E_{12}^2+E_p^2}$, where $E_p$ is the plasma energy. The energy $\tilde{E}_{12}$ corresponds to the limit of the energy of the upper polariton for a cavity width infinitely large. It is important to notice that in our devices the cavity modes do not depend on the incidence angle $\theta$, as the modes confined under the metal strips are decoupled~\cite{todorov_opex} and give rise to a flat polariton dispersion with respect to the component of the photon wavevector perpendicular to the strips, $k_x$. This is shown in figure~\ref{reflectivity}b, where the reflectivity minima measured at 300~K on the same sample HM2952 at resonance are shown as a function of $k_x$ (white dots). The experimental points are compared with the simulated reflectivity, which is presented in a grey scale. In the simulation, the quantum well layers are modeled within the random phase approximation by a diagonal dielectric tensor, as in Ref.~\citenum{Wendler_Kandler}, while for the gold dielectric constant we used the same values as in Ref.~\citenum{jouy_apl}.

\begin{figure}[ht]
\centering
\includegraphics[width=0.8\columnwidth]{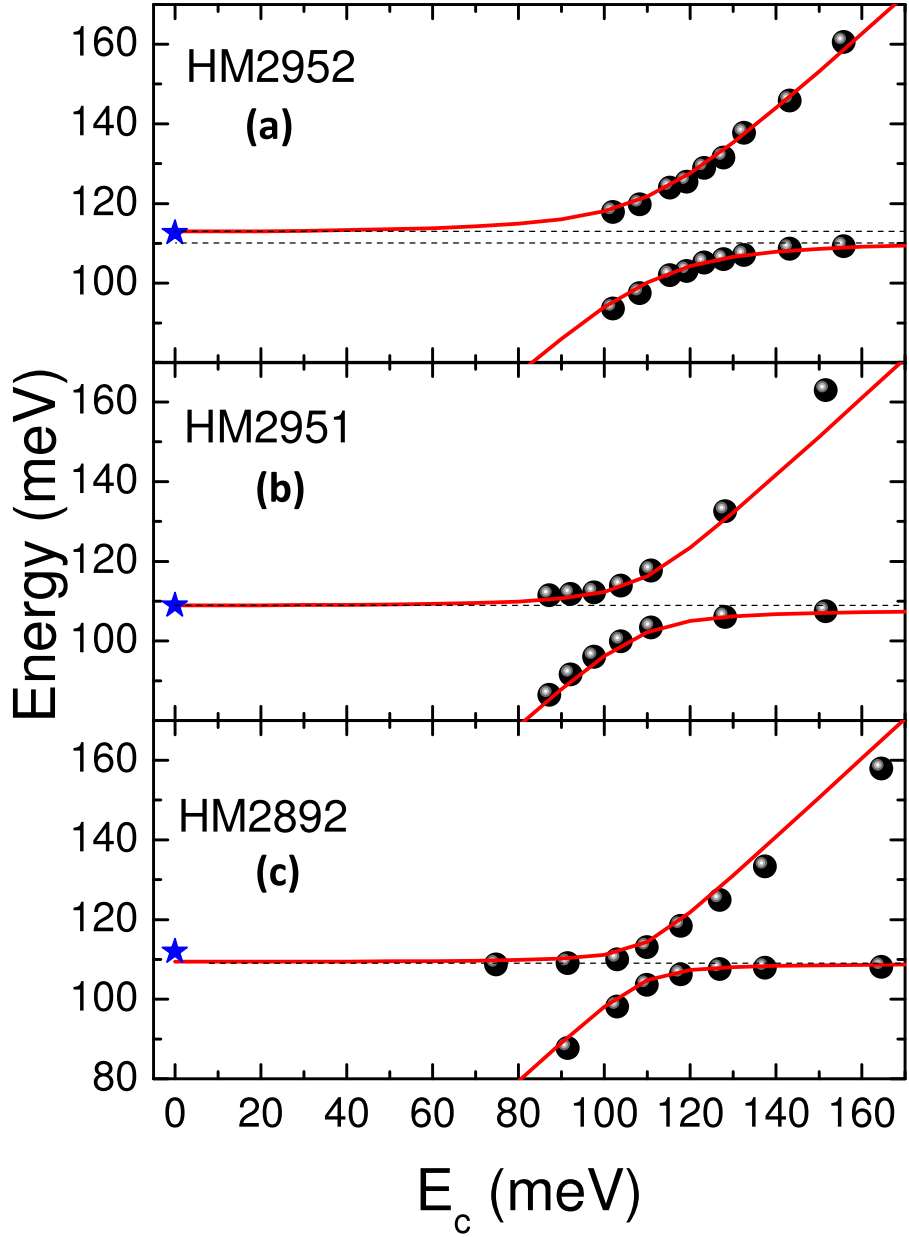}
\caption{Reflectivity minima at 300~K (bullets), plotted as function of the cavity mode energy. The continuous line is the simulated polariton dispersion, following eq.~\ref{ultra_strong}. The dashed horizontal lines are the asymptotes of the polariton branches. The star indicates the ISB absorption measured from transmission experiments at 300~K on an unprocessed sample.}
\label{dispersion}
\end{figure}

\begin{figure}[ht]
\centering
\includegraphics[width=0.8\columnwidth]{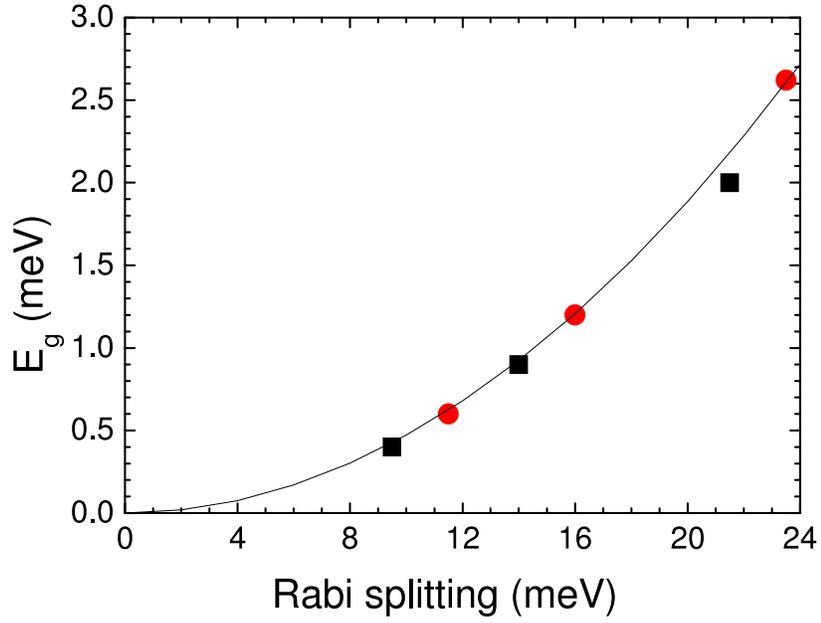}
\caption{Energy gap extracted from the measured dispersion of the three samples plotted as a function of the Rabi splitting at 300~K (squares) and 77~K (circles). The continuous line represents eq.~\ref{energy_gap}, with $E_{12}=106$~meV. }
\label{gap}
\end{figure}

The polariton dispersion extracted from the reflectivity spectra is shown in figure~\ref{dispersion}a, where the minima (bullets) are plotted as function of the cavity mode energy. The minimum splitting between the upper and lower branch, $2 \hbar \Omega_R$, is 21.5~meV. Thanks to the very high confinement of the light in the MDM cavity, we obtain at room temperature a ratio $\hbar \Omega_R/E_{12}=0.1$, with a system containing only 10 quantum wells. This value is comparable to that obtained in Ref.~\citenum{anappara_PRB} using a planar microcavity with 70 quantum wells with the same doping level as in our sample.
The polariton dispersion curve can be calculated by solving the following eigenvalue equation~\cite{todorov_PRL2010}:
\begin{equation}
\left( E^2 - \tilde{E}_{12}^2\right) \left(E^2 -E_c^2 \right)=4 \left(\hbar \Omega_R \right)^2 \, E_c^2
\label{ultra_strong}
\end{equation}
where $2 \hbar \Omega_R=\sqrt{f_w} E_p$, $f_w=0.35$ is the overlap factor between the quantum wells and the cavity mode and $E_c$ is the energy of the cavity mode. Equation~\ref{ultra_strong} provides the eigenvalues of the complete light-matter interaction Hamiltonian, including the quadratic term, which accounts for the polarization self-interaction, and the antiresonant terms, necessary to describe a system in the ultra-strong coupling regime~\cite{ciuti_PRB2005}. In this case the minimum energy splitting $2 \hbar \Omega_R$  does not occur at resonance, $E_c = E_{12}$, but when the energy of the cavity mode is equal to $E_{\rm{min}}= E_{12} \, \sqrt{1+\frac{E_p^2}{E_{12}^2}\left( 1-f_w \right)}$. Interestingly this value corresponds also to that of the asymptote of the lower branch for $E_c \gg E_{12}$. In the opposite limit, for $E_c \ll E_{12}$, as already mentioned the upper polariton branch tends to $\tilde{E}_{12}$. In Fig.~\ref{dispersion} we indicated with a star the renormalized ISB absorption energy as obtained from transmission experiments on the corresponding unprocessed sample (see dashed line in Fig.~\ref{reflectivity}). Notice that they are in excellent agreement with the asymptotic value of the polariton for $E_c$ going towards zero. It is important to remark that the parameters entering in eq.~\ref{ultra_strong} are determined from experimental data. The plasma energy is inferred from the value of the minimum splitting, while the ISB transition energy is extracted from the value $E_{\rm{min}}$ of the asymptote of the lower polariton dispersion. Moreover, the value of $\tilde{E}_{12}$ can be consistently checked by using the upper polariton horizontal asymptote. The value that we found for $E_{12}$ is very close to that calculated using Schr\"{o}dinger equation in the Hartree approximation. From the plasma energy, we extract a value of $8.5 \times 10^{11} \, \rm{cm}^{-2}$ for the difference between the electronic populations of the two subbands, $N_1-N_2$. This value is lower than the one expected in our sample at 300~K ($2.3 \times 10^{12} \, \rm{cm}^{-2}$). This discrepancy was already observed in the literature~\cite{geiser} and it can be due to the localization of a fraction of the electrons in defects originated from the wafer bonding process. The calculated polariton dispersion, using these parameters and eq.~\ref{ultra_strong}, is shown in Fig.~\ref{dispersion}a by a continuous line and reproduces very well the experimental data including the energy of the measured ISB absorption.
Figures~\ref{dispersion}b and \ref{dispersion}c presents the reflectivity minima (bullets) and the polaritonic dispersion calculated using eq.~\ref{ultra_strong} respectively for sample HM2951 and HM2892. As in fig.~\ref{dispersion}a, we reported with a star the value of the ISB absorption energy extracted from room temperature absorption experiments on corresponding unprocessed samples (labeled as ISB abs. in Table~\ref{tabella}). Table~\ref{tabella} reports the values of the plasma energy and of the ISB transition energy extracted from polariton dispersions. In the table we also included the results obtained at 77~K on the three samples, where we have observed a small increase of the value of the Rabi splitting, which is consistent with an increase of the electronic population on the ground subband.
Figure~\ref{dispersion}a shows that the polaritonic dispersion presents a photonic gap, delimited by two dashed lines, which are the horizontal asymptotes of the polariton branches. As discussed in ref.~\citenum{todorov_PRL2010}, this gap directly results from eq.~\ref{ultra_strong} and has its origins in the quadratic and anti-resonant terms of the light-matter interaction Hamiltonian. As a consequence, the observation of a polaritonic gap in the dispersion curve can be taken as the experimental signature of the ultra-strong coupling regime. The value of the energy gap is:
\begin{equation}
E_g \approx f_w \left( \tilde{E}_{12} -E_{12}\right) = \left( 2 \hbar \Omega_R \right)^2 / \left( 2 E_{12}\right)
\label{energy_gap}
\end{equation}
The right side is a development up to the third order in $E_P/E_{12}$. From this expression it is evident that the ratio $E_g/E_{12}$ has a quadratic dependence on $\Omega_R/E_{12}$ and therefore appears only in the regime of ultra-strong coupling. Fig.~\ref{gap} presents the values of $E_g$ for the three samples, both at 300~K (squares) and 77~K (circles), extracted from the polaritonic dispersions, and plotted as a function of the measured Rabi splitting. The continuous line gives the quadratic dependence on the Rabi energy, by using $E_{12}=106$~meV. The left hand side of eq.~\ref{energy_gap} clearly shows the two quantities responsible for the opening of the photonic gap, thus for the entering in the ultra-strong coupling regime. To maximize the gap one has therefore to optimize $f_w$ and increase the difference between the ISB transition energy and its renormalized value due to the collective effect of the two-dimensional electron gas. In samples HM2951 and HM2892 the gap is not visible in the polariton dispersion. In this case the quadratic and anti-resonant terms can be neglected in the light-matter interaction Hamiltonian and the polariton dispersion is well simulated by solving the usual strong coupling equation: $\left( E-E_{12}\right) \left( E-E_c\right)=\left( \hbar \Omega_R \right)^2$.

In conclusion, we demonstrated the achievement of the ultrastrong coupling regime between a mid-infrared ISB transition ($\lambda =11.7~\mu$m) and the fundamental mode of a MDM microcavity. The experimental signature of this regime is the opening of a photonic gap in the polariton dispersion. The transition from the strong to the ultra-strong coupling regime has been investigated by comparing the dispersions of three identical samples with different doping levels. As a perspective, MDM cavities could be employed for realizing active devices operating in the ultra-strong coupling regime. In this case, thanks to the large value of the splitting, a selective electrical injection into the polariton states could be achieved, avoiding the excitation of the dark states at the bare ISB transition energy, which dominate the transport in existing intersubband polariton devices.
\begin{acknowledgments}
We thank M. Rosticher for help with the e-beam lithography system. This work has been partially supported by the French National Research Agency (ANR) in the frame of its Nanotechnology and Nanosystems program P2N, Project No. ANR-09-NANO-007. We acknowledge financial support from the ERC grant ``ADEQUATE''.
\end{acknowledgments}

\newpage

\newpage
\begin{table} \caption{Relevant quantities for the three samples at 300~K (77~K).}
\begin{tabular}{Sc Sc Sc Sc Sc Sc Sc}
\hline
\hline
   & $2 \hbar \Omega_R$ (meV) & $E_P$ (meV) & $E_{12}$ (meV) & $N_1-N_2$ ($\times 10^{11} \rm{cm}^{-2}$) & $\tilde{E}_{12}$ (meV) & ISB abs. (meV) \\
   \hline
HM2952 & 21.5 (23.5) & 36.5 (39.9) & 107 (105.2) & 8.5 (10) & 113 (112.5) & 112.5 \\
 HM2951 & 14 (16) & 23.8 (27.2) & 106.3 (108.3) & 3.6 (4.7) & 108.9 (111.7) & 108.9 \\
 HM2892 & 9.5 (11.5) & 16.1 (19.5) & 108.2 (106.8) & 1.7 (2.4) & 109.4 (108.6) & 112\\
\hline
\hline
\end{tabular}
\label{tabella}
\end{table}

\end{document}